\documentclass[12pt]{article} 
\usepackage[utf8x]{inputenc} 
\usepackage[T1]{fontenc} 
\usepackage{amsmath,amssymb,bbm,bm,commath,xfrac} 
\usepackage{graphicx,subcaption} 
\graphicspath{{figures/}}
\usepackage{color} 
\usepackage{cite} 

\usepackage{geometry} 
\usepackage{fancyhdr} 
\usepackage{setspace} 
\usepackage[nottoc,notlof,notlot]{tocbibind} 

\usepackage[unicode]{hyperref} 
\hypersetup{bookmarksnumbered=true, bookmarksopen=true, bookmarksopenlevel=2, breaklinks=true, citecolor=blue, colorlinks=true, linkcolor=red, linktoc=page, pdfborder={0 0 0}, pdfstartview=FitH, plainpages=false, unicode=true, urlcolor=blue}

\DeclareUnicodeCharacter{"0393}{\Gamma}
\DeclareUnicodeCharacter{"0394}{\Delta}
\DeclareUnicodeCharacter{"0398}{\Theta}
\DeclareUnicodeCharacter{"039B}{\Lambda}
\DeclareUnicodeCharacter{"039E}{\Xi}
\DeclareUnicodeCharacter{"03A0}{\Pi}
\DeclareUnicodeCharacter{"03A3}{\Sigma}
\DeclareUnicodeCharacter{"03A5}{\Upsilon}
\DeclareUnicodeCharacter{"03A6}{\Phi}
\DeclareUnicodeCharacter{"03A8}{\Psi}
\DeclareUnicodeCharacter{"03A9}{\Omega}
\DeclareUnicodeCharacter{"03B1}{\alpha}
\DeclareUnicodeCharacter{"03B2}{\beta}
\DeclareUnicodeCharacter{"03B3}{\gamma}
\DeclareUnicodeCharacter{"03B4}{\delta}
\DeclareUnicodeCharacter{"03B5}{\epsilon}
\DeclareUnicodeCharacter{"03B6}{\zeta}
\DeclareUnicodeCharacter{"03B7}{\eta}
\DeclareUnicodeCharacter{"03B8}{\theta}
\DeclareUnicodeCharacter{"03D1}{\vartheta}
\DeclareUnicodeCharacter{"03B9}{\iota}
\DeclareUnicodeCharacter{"03BA}{\kappa}
\DeclareUnicodeCharacter{"03BB}{\lambda}
\DeclareUnicodeCharacter{"03BC}{\mu}
\DeclareUnicodeCharacter{"03BD}{\nu}
\DeclareUnicodeCharacter{"03BE}{\xi}
\DeclareUnicodeCharacter{"03C0}{\pi}
\DeclareUnicodeCharacter{"03C1}{\rho}
\DeclareUnicodeCharacter{"03C3}{\sigma} 
\DeclareUnicodeCharacter{"03C4}{\tau}
\DeclareUnicodeCharacter{"03C5}{\upsilon}
\DeclareUnicodeCharacter{"03C6}{\phi}
\DeclareUnicodeCharacter{"03D5}{\varphi}
\DeclareUnicodeCharacter{"03C7}{\chi}
\DeclareUnicodeCharacter{"03C8}{\psi}
\DeclareUnicodeCharacter{"03C9}{\omega}
\DeclareUnicodeCharacter{"21D0}{\Leftarrow}
\DeclareUnicodeCharacter{"0212B}{\AA}
\DeclareUnicodeCharacter{"00B7}{\cdot}
\DeclareUnicodeCharacter{"00B0}{^{\circ}}
\DeclareUnicodeCharacter{"266A}{\eighthnote}
\DeclareUnicodeCharacter{"266B}{\twonotes}
\PrerenderUnicode{ä}\PrerenderUnicode{×}

\definecolor{title}{rgb}{0,0,0}
\definecolor{abst}{rgb}{0,0,0}
\definecolor{sect}{rgb}{0,0,0}
\definecolor{ssect}{rgb}{0,0,0}
\definecolor{sssect}{rgb}{0,0,0}
\definecolor{appsect}{rgb}{0,0,0}
\definecolor{ref}{rgb}{0,0,0}
\newcommand{\Title}[1] {\title{\color{title}\Huge #1}}
\newcommand{\TPheader}[3] {\date{}\maketitle\thispagestyle{fancy}\pagenumbering{alph}\lhead{#1}\chead{#2}\rhead{#3}\cfoot{}}
\newcommand{\makepage}[1] {\newpage\pagenumbering{#1}}

\newcommand{\Abstract}[1] {\begin{abstract}\normalsize #1 \end{abstract}}

\renewcommand{\appendix}{\setcounter{section}{0}\renewcommand{\thesection}{\Alph{section}}\renewcommand*{\theHsection}{app.\the\value{section}}} 
\newcommand\references[1]{\bibliographystyle{hephys}\bibliography{#1}}

\newcommand\eqs[1] {\begin{align}#1\end{align}}

\newcommand\eqsc[1] {\begin{gather}#1\end{gather}}

\newcommand\equ[1] {\begin{equation}#1\end{equation}}

\newcommand\fig[2] {\begin{figure}[#1]\centering #2\end{figure}}

\newcommand\s {\sigma}

\renewcommand\( {\left(}
\renewcommand\) {\right)}

\newcommand\A {{\mathcal A}}
\renewcommand\C {{\mathcal C}}

\newcommand\E {{\mathcal E}}
\renewcommand\G {{\mathcal G}}

\newcommand\N {{\mathcal N}}

\newcommand\ie {\textit{i.e.}}
\newcommand\eg {\textit{e.g.}}
\newcommand\etc {\textit{etc.}}
\newcommand\nn {\nonumber\\}

\numberwithin{equation}{section} 
\begin{document}
\Title{Rhombic Patterns Near a Bicritical Point in Periodically Forced Surface Waves}

\author{Krishna Kumar\footnote{\href{mailto:kumar@phy.iitkgp.ernet.in}{kumar@phy.iitkgp.ernet.in}}$~$, Supriyo Paul and Dharmesh Jain \bigskip\\
\emph{\normalsize Department of Physics and Meteorology, Indian Institute of Technology,}\\
\emph{\normalsize Kharagpur-721302, West Bengal, India}
}

\TPheader{}{{\it Indian J. Phys.} {\bf 81[11]} (2007) 1205}{} 

\Abstract{We present here a study of selection of rhombic patterns close to a bicritical point at the onset of primary surface instability in viscous fluids under two-frequency vertical vibration. Rhombic patterns appear to be natural at the primary instability in the form of a bicritical point if the ratio of driving frequencies is selected properly. We present two different patterns which may be accessible in a Faraday experiment.\\


}

\tableofcontents
\makepage{arabic}

\section{Introduction}
Forced extended dissipative systems often show generation patterns when driven far from thermodynamic equilibrium. These patterns, which are generated spontaneously, break some (at least one) symmetries of the system in the old state. These patterns, also known as \emph{dissipative structures}, can be maintained in mechanical equilibrium by controlling the external drive. The examples include formation of convective cells in Rayleigh-Bénard convection \cite{Rayleigh:1883df,Benard:1901hf} and Couette-Taylor \cite{Couette:1890:mf,Taylor:1923gi} flow, surface waves in Faraday experiment \cite{Faraday:1831mf}, magnetic fluids \cite{Bashtovoi:1993mm}, liquid crystals \cite{Alexander:1990low,Murray:1991cg}, {\etc} Recent works on parametric excitation of waves in fluids address various physical problems such as forced surface waves \cite{Perlin:2000ws}, thermo-capillary waves \cite{Skarda:2001jrl,Or:2002rek}, pattern forming instabilities \cite{Fauve:1992kk,Edwards:1993sf,Muller:1993hw,Edwards:1994sf,Kumar:1995kb,Kudrolli:1996jpg,Binks:1997wvw,Zhang:1997jv,Chen:1997jv,Silber:1998mp,Kudrolli:1998pg,Arbell:2000jf,Porter:2002ms,Wagner:2003mk}, multicritical points \cite{Kumar:1996lf,Besson:1996et,Silber:1999as,Kumar:2004bm} at the onset of Faraday waves, {\etc} The experimental arrangement in Faraday set up has attracted considerable attention in recent years due to its simplicity and its ability to excite a plethora of patterns. The possibility of various kinds of bicritical points \cite{Kumar:1996lf,Besson:1996et,Silber:1999as,Kumar:2004bm}, which appear in thin layers of viscous fluids, may lead to exotic patterns \cite{Muller:1993hw,Edwards:1994sf,Kumar:1995kb,Kudrolli:1998pg,Silber:1998mp,Porter:2002ms} at the onset of Faraday instability. Bicritical points, where two standing waves of different spatial and temporal periodicity may be excited simultaneously, can be forced by using sinusoidal \cite{Kumar:1996lf,Kumar:2004bm} as well as multi-frequency forcing \cite{Besson:1996et}. Linear stability analysis of parametrically forced surface waves with two frequency forcing was first done by Besson, et al \cite{Besson:1996et} for a chosen frequency ratio. Later, Silber and Skeldon \cite{Silber:1999as} analyzed many frequency ratios and considered some specific patterns. Arbell and Fineberg \cite{Arbell:2000jf} did experiments with two frequency forcing showing several exotic patterns. However, there is no systematic study on the role of two different forcing frequencies on bicritical points. It is not known how to predict the excited tongues once two frequencies are selected.

\

The selection of patterns is usually described by amplitude equations. The normal form of these equations depends on the broken symmetries of the old state. The normal form may depend on the dissipation in the system even if the same symmetry is broken. The amplitude equations provide a common framework to analyze patterns in the close vicinity of an instability in a system or a bifurcation of a solution. Parametrically forced dissipative structures provide another class of pattern-forming systems. Pattern selection in such systems may be controlled to some extent by the external drive. The spatio-temporal patterns follow some dispersion relation at the onset. The period of the patterns and therefore their size may be controlled, if the system may be forced to resonate with the external driving. This is what precisely happens in a parametrically driven system. Faraday waves on fluid surface and Langmuir waves in plasma are well known examples of such cases.

\

In this paper, we investigate selection of primary patterns due to spontaneous excitation of surface waves just above the onset of a bicritical point in a fluid vibrated vertically with two frequencies. The two frequencies may be chosen as $lω$ and $mω$, where $ω$ is a basic frequency. The integers $l$ and $m$ may be \emph{odd-odd}, \emph{odd-even} or \emph{even-odd} combinations. The \emph{even-even} combination would reduce to one of the cases mentioned above. Rhombic patterns appear easily accessible, if we choose odd component smaller than the even component. Many other complex and exotic patterns may be possible just at the instability onset in this case. In other cases, secondary or tertiary patterns instead of the primary pattern are likely to be more complex.

\section{Hydrodynamic System}
We consider a laterally extended layer of incompressible Boussinesq fluid of kinematic viscosity $ν$, thermal diffusivity $κ$, and thickness $h$ resting on a flat and rigid plate which is subjected to a vertical oscillation of the form: $F(t)= ε\(\cos χ \cos(lωt) +\sin χ\cos(mωt + φ)\)$, where $χ$ is mixing angle and $ω$ is the basic frequency. In a frame of reference fixed with the oscillating plate, the free surface of the fluid is initially flat, stationary, and coincident with $z=0$ plane. The oscillation is then equivalent to a temporally modulated gravitational acceleration $G(t)=g -F(t)$. The basic state of rest has a time dependent pressure $P(t)= P_0 -ρG(t)z$, where $P_0$ is the uniform atmospheric pressure and $ρ$ is the uniform reference density of the fluid. For the linear stability analysis, all the nonlinear terms are dropped. We may also compute the fields and their derivatives at the free-surface by Taylor-expanding them about the flat free-surface $(z=0)$. Lengths, time and pressure are re-scaled by the original fluid thickness $h$, viscous diffusion time $h^2/ν$ and $ρν^2/h^2$, respectively. The linearized version of the dimensionless hydrodynamic equations in the bulk then read
\eqs{∂_t\bm{v} &=-∇p +∇^2\bm{v}\,, \label{eq1}\\
∇·\bm{v} &=0\,.
}
Notice that there is no term due to the buoyancy force in the momentum equation \eqref{eq1}. This is the case in Boussinesq fluids in the presence of small surface deformation. The surface tension is measured by the inverse of the capillary number $C= ρν^2/\s_0h$. The fluid rests on a rigid plate at which the velocity field must vanish. At the onset of surface waves, the free surface is located at $z= z(\bm{x},t)$ where $\bm{x}= (x_1,x_2)$ lies in the horizontal plane, and obeys the kinematic condition \cite{Lamb:1932cup} given as $∂_tζ =v_3|_{z=0}$. The dynamic boundary conditions are determined by the stress tensor $Π_{ij}= -pδ_{ij} +(∂_iv_j +∂_jv_i) +\G(t)ζδ_{ij}$ at the free surface. The jump in the normal stress at any point on the free surface is equal to the surface tension $Σ$ times the curvature at that point. But every component of the tangential stress vanishes everywhere on the free surface. For the linearized system, they are
\eqs{\left[\(∂_t -∂_{zz} -3∇_H^2\)∂_zv_3\right]_{z=0} &=\left[\G - \C^{-1}∇_H^2 -\E\left\{\cos χ\cos(lΩt) +\sin χ\cos(mΩt + φ)\right\}\right]∇_H^2ζ\,, \\
\left[\(∂_{zz} -∇_H^2\)v_3\right]_{z=0} &=0\,.
}
where $\E =ε h^3/ν^2$ the dimensionless forcing amplitude and $Ω =ωh^2/ν$ the dimensionless forcing frequency. The dependence on horizontal coordinates of all the fields may be expressed in terms of the normal modes $\sin(\bm{k}·\bm{x})$ of the horizontal plane, with the horizontal wave number $k =\sqrt{k_1^2 +k_2^2}$ and position vector $\bm{x}$ in horizontal plane. The linear stability problem then explicitly depends only on the vertical coordinate $z$ and time $t$. The stability problem is analyzed by Floquet theory. We expand the fields as
\eqs{v_3(z,t) &=e^{μt} ∑_{n=-∞}^{∞} w_n(z)\,e^{i n Ω t}\,, \nn
ζ(t) &=e^{μt} ∑_{n=-∞}^{∞} ζ_n\,e^{i n Ω t}\,.
}
The Floquet exponent $μ =s +iαΩ$ is a complex number, where $s$ and $α$ are real and finite. The solutions corresponding to $α=0$ and $α=\frac{1}{2}$ are referred to as harmonic and subharmonic solutions, respectively. The relevant equations describing the complete linear stability of the horizontally infinite layer are then given by
\eqsc{\(D^2 -k^2\)\(D^2 -q_n^2\)w_n =0\,, \label{eq6} \\
w_n =Dw_n =0 \quad \text{ at } z=-1\,, \label{eq7} \\
\left[s+i(α+n)Ω\right]ζ_n =w_n \quad \text{ at } z=0\,. \label{eq8}
}
The stress conditions at $z=0$ are:
\eqs{\(D^2+k^2\)w_n &=0\,, \label{eq9} \\
\(D^2 -q_n^2 -2k^2\)Dw_n  -\(\G +C^{-1}k^2\)k^2ζ_n &= -\E k^2\big[\cos χ\(ζ_{n-l} +ζ_{n+l}\) \nn
&\qquad\qquad +\sin χ\big(e^{iφ}ζ_{n-m} +e^{-iφ}ζ_{n+m}\big)\big]\,, \label{eq10}
}
where $D≡∂_z$ and $q_n^2 =k^2 +[s +i(α+n)Ω]$. The general solution of the bulk equations \eqref{eq6}, may be written as,
\equ{w_n =P_n\cosh(kz) +Q_n\sinh(kz) +R_n\cosh(q_nz) +S_n\sinh(q_nz)\,,
\label{eq11}}
applying the boundary conditions \eqref{eq7}-\eqref{eq9} on the general solution \eqref{eq11} we obtain the four unknowns $P_n$, $Q_n$, $R_n$ and $S_n$ in terms of $ζ_n$ for every $n$:
\eqs{P_n &= \big(k^2 +q_n^2\big)ζ_n\,, \\
R_n &= -2k^2ζ_n\,, \\
S_n &= -\frac{k P_n +R_n\(k\cosh q_n \cosh k -q_n \sinh q_n \sinh k\)}{q_n \cosh q_n \sinh k -k \sinh q_n \cosh k}\,, \\
Q_n &= \frac{q_n R_n +P_n\(q_n\cosh q_n \cosh k -k \sinh q_n \sinh k\)}{q_n \cosh q_n \sinh k -k \sinh q_n \cosh k}\,·
}
Now applying the condition \eqref{eq10} for the pressure jump across the free surface leads to a recursion relation of the form:
\equ{A_nζ_n =\E\big[\cos χ\(ζ_{n-l} +ζ_{n+l}\) +\sin χ\big(e^{iφ}ζ_{n-m} +e^{-iφ}ζ_{n+m}\big)\big]\,,
\label{eq16}}
where
\equ{A_n=2\left[\G +\frac{k^2}{\C} -\frac{4 q_n k^2(q_n^2 +k^2) -C_n \cosh q_n \cosh k +D_n\sinh q_n \sinh k}{k\(q_n \cosh q_n \sinh k -k \sinh q_n \cosh k\)}\right],
}
with $C_n =q_n\(q_n^4 +2q_n^2k^2+ 5k^4\)$ and $D_n=k\(q_n^4 +6q_n^2k^2 +k^4\)$. Also, the boundary condition \eqref{eq8} and continuity conditions ensures that $w_0(z)=0$ for all $z$ when $(μ +inΩ)=0$. Therefore
\equ{A_0(μ +inΩ=0) ≡A_0^h =2\left[\G +\C^{-2}k^2\right].
}
The stability of the free surface can be determined numerically with any preassigned accuracy by converting the recursion relations \eqref{eq16} to an ordinary eigenvalue problem \cite{Kumar:1996lf,Kumar:2004bm} given as $(\A^{-1}\N)\bm{Z} = \E^{-1}\bm{Z}$. The marginal stability boundaries are defined by the curves in $\E -k$ plane on which $s(\E,k)=0$.

Figure \ref{fig1} shows the marginal stability curves for two-frequency forcing in glycerol-water mixture. The zones bounded by continuous line are the regions of the $(\E/\G -k)$ where the excited standing waves are temporally subharmonic with respect to the forcing. When the lowest points of two different tongues have the same forcing amplitude, we have a bicritical point as primary instability. Two different wave numbers are then excited simultaneously at the instability onset. We denote the first number of the pair of integers by $l$ and the other by $m$. The panel of four plots (clockwise from the top left) in Figure \ref{fig1} show bicritical points for the frequency ratio $l:m$ equal to $5:8$, $8:13$, $13:21$ and $21:34$, respectively. This combination is responsible for the selection of wave numbers. For \emph{odd-odd} combination (see bottom right) we have two wave numbers belonging to the lowest point of two subharmonic tongues. The $i^{th}$ and $j^{th}$ subharmonic tongues with $i=(l+1)/2$ and $j=(m+1)/2$ are excited. For \emph{even-odd} combinations (top right), the bicritical points involve one harmonic and other subharmonic solutions. If $l$ is odd and $m$ even (as in the left column), $i^{th}$ with $i=(l+1)/2$ subharmonic tongue and $j^{th}$ with $j=m/2$ harmonic tongues are excited at the onset of the bicritical point. Similarly, for $l$ even and $m$ odd, $i^{th}$ with $i=l/2$ harmonic and $j^{th}$ with $j=(m+1)/2$ subharmonic tongues are excited at the onset. If the ratio $l:m$ is kept fixed, the same tongues depending on the above three combinations, would be excited. This rule appears to be always correct except in thin layers of viscous fluids or at very low excitation frequencies.
\vspace*{-6mm}
\fig{!h}{\includegraphics[scale=1]{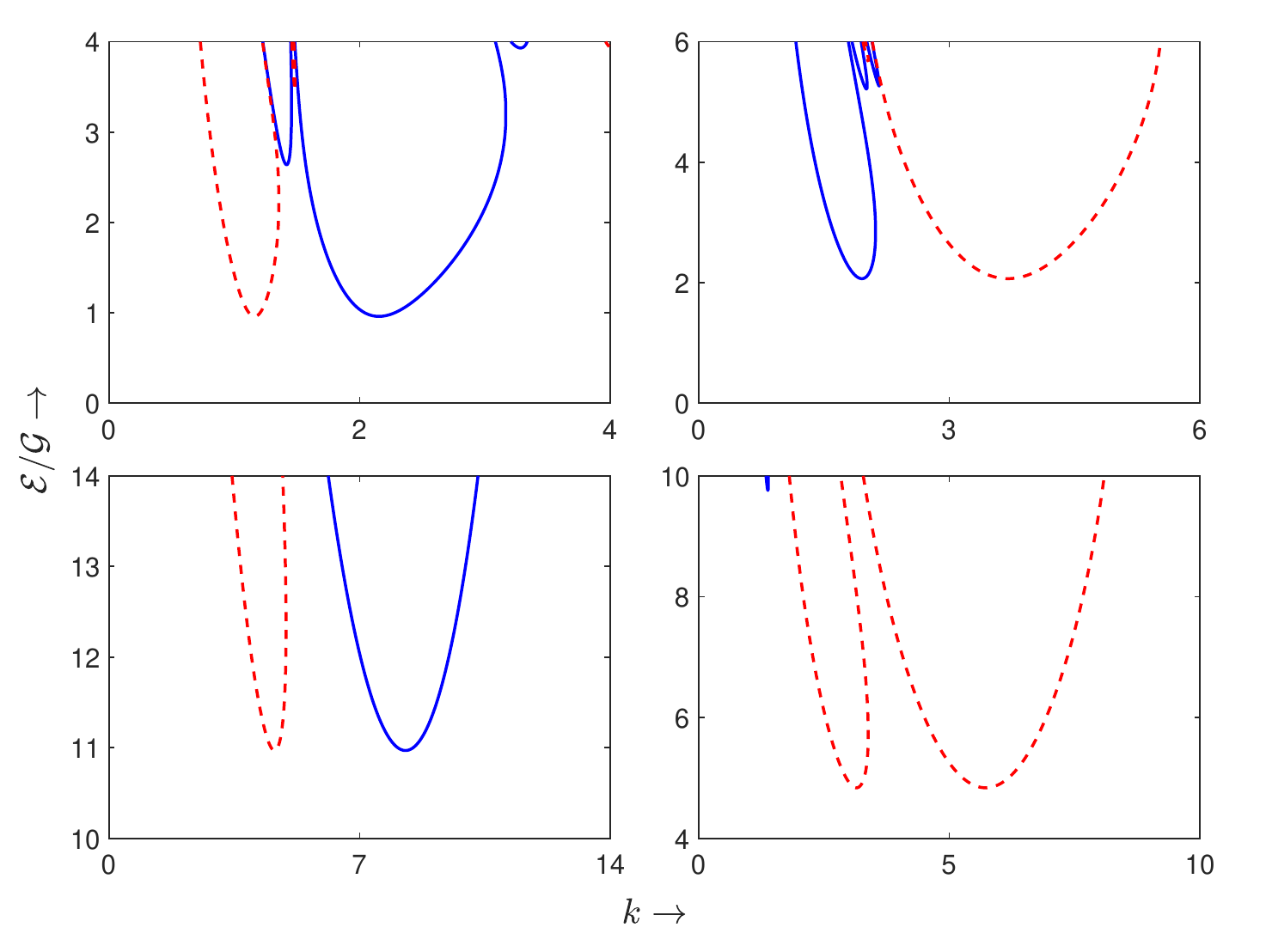}
\caption{Stability boundaries for subharmonic (regions bounded by red, dashed line) and harmonic (regions bounded by blue, solid line) solutions in glycerol-water mixture. The tongue-like zones in panel of four figures (clockwise from top left) are for frequency ratios ($l:m$) equal to $5:8$, $8:13$, $13:21$ and $21:34$, respectively. Other fluid parameters are: $\G=9.42×10^2$, $Ω=12.32$, $\C=1.74×10^{-2}$ and $h=1$ cm.}
\label{fig1}}

Figure \ref{fig2} displays the temporal part of the critical mode for surface deformation for 1 cm thick layer of glycerol-water mixture for frequency ratio $l:m=5:8$ (upper row) and $8:13$ (lower row). The bicritical point for frequency ratio $5:8$ involves two tongues: third subharmonic (SH3) and fourth harmonic (H4). The temporal parts of the critical deformation $ζ_c$ corresponding to the frequency $5:8$ (upper row) show five and eight maxima/minima, respectively, in a period equal to double the forcing period $T=2π/Ω$. The temporal parts of the critical deformation for frequency ratio $8:13$ (lower row) show eight and thirteen maxima/minima in a period equal to $2T$. The critical modes with periodicity equal to $2T$ are labeled `SH', and those with periodicity $T$ are labeled `H'. The third subharmonic tongue has dominant frequency $(2×3 -1)Ω/2=5(Ω/2)$ at its lowest point of the tongue. So, the critical deformation showing five maxima/minima in a period of $2T$ is designated as `SH3'. Similarly, fourth harmonic tongue has dominant frequency $4Ω=8(Ω/2)$ at its lowest point. So, the critical deformation corresponding to dominant frequency $8Ω$ in two frequency forcing shows four maxima/minima in a period equal to $2T$. The presence of other frequencies at the onset of surface waves distorts the critical mode but the number of local maxima/minima at the instability onset is governed by the dominant frequencies imposed.
\fig{!h}{\includegraphics[scale=1]{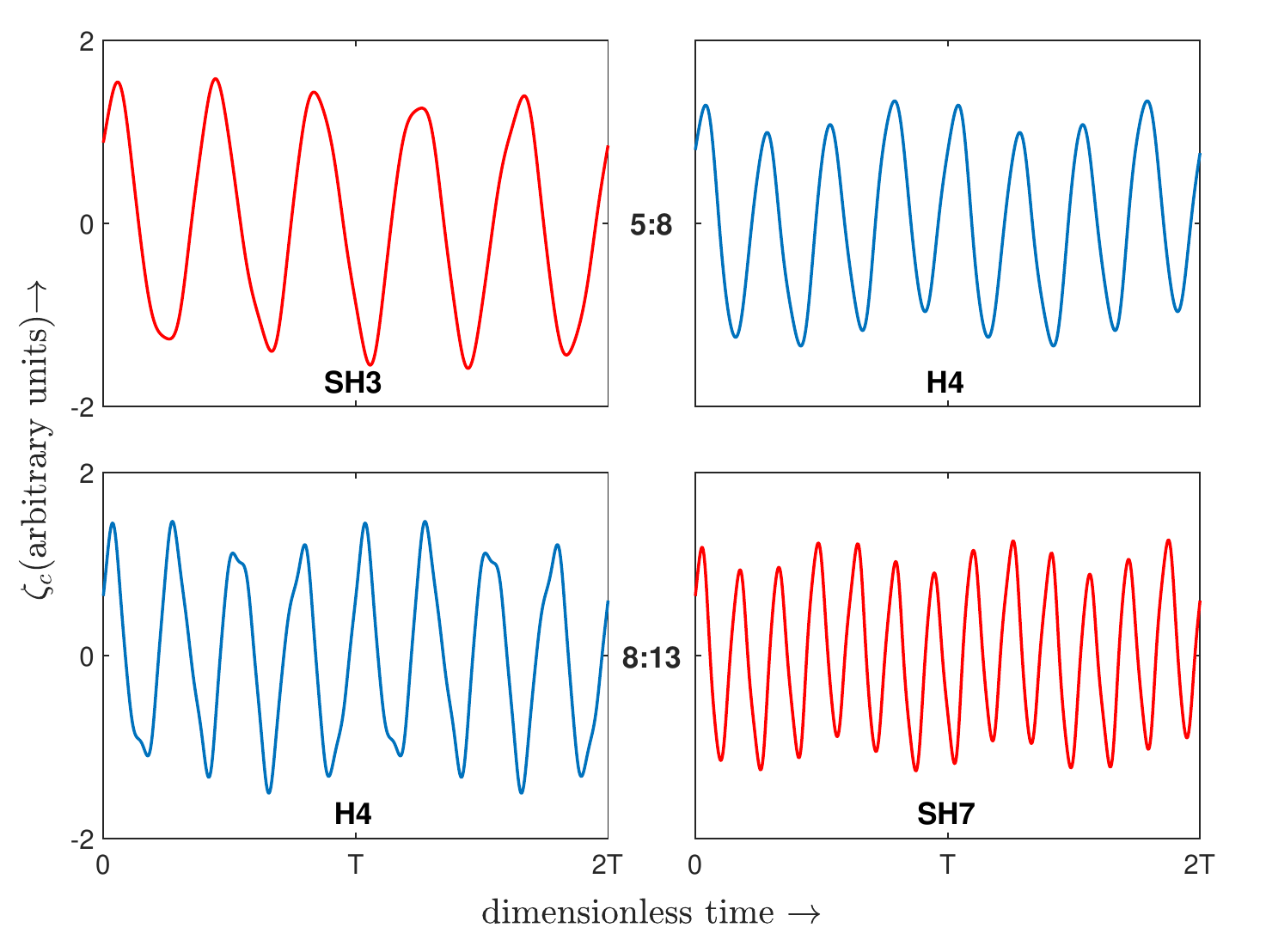}
\caption{Temporal parts of the critical modes $ζ_c$ for surface deformation at the bicritical point of 1 cm thick layer of glycerol-water mixture for frequency ratios $5:8$ (upper row) and $8:13$ (lower row) with parameters same as those given in Figure \ref{fig1}.}
\label{fig2}}

\section{Triad Interaction}
Formation of surface wave patterns involves interaction between standing waves of different wave vectors. The modes form a resonant triad \cite{Silber:1999as} if the wavevectors $\bm{k_1}$, $\bm{k_2}$ and $\bm{k_3}$ satisfy,
\equ{\bm{k_1} +\bm{k_2} =\bm{k_3}
}
with $|\bm{k_1}| =|\bm{k_2}| =k_s$, and $|\bm{k_3}| =k_h$ along with the frequencies $ω_1 +ω_2 =ω_3$. $\bm{k_1}(ω_1)$ and $\bm{k_2}(ω_2)$ are the wavevectors (frequency) corresponding to the first tongue and $\bm{k_3}(ω_3)$ is that of second tongue at a bicritical point. Hence, we can write the resonant triad as,
\equ{ζ(t) =A_1(T)e^{i\bm{k_1}·\bm{x}} e^{iω_1t} +A_2(T)e^{i\bm{k_2}·\bm{x}} e^{iω_2t} + B(T)e^{i\bm{k_3}·\bm{x}} e^{iω_3t} +⋯ +c.c.\,,
\label{eq21}}
where $A_j$ (with $j=1,2$) and $B$ are the complex amplitudes of the first and second tongue, respectively, at the bicritical point and $T$ is slow time scale. The periodic function $ζ(t)$ has a time period of $2π/Ω$. Considering the symmetries we obtain the coupled amplitude equations for the resonant triad:
\eqs{\dot{A}_1 &=ε A_1 +μA_2^* B -\left[α\(|A_1|^2 +|A_2|^2\) +β|B|^2\right] A_1\,, \nn
\dot{A}_2 &=ε A_2 +μA_1^* B -\left[α\(|A_1|^2 +|A_2|^2\) +β|B|^2\right] A_2\,, \nn
\dot{B} &=ε B +ν A_1 A_2 -\left[δ\(|A_1|^2 +|A_2|^2\) +γ|B|^2\right]B\,,
\label{eq22}}
where the coefficients are all real. For simplicity we have assumed that the interaction coefficients between two modes of a single tongue are same. It is to be noted that there is another set of equations which are complex conjugate to the equations \eqref{eq22}. The general solutions of the amplitude equations are complex and can be written as,
\equ{A_1 =R_1e^{iθ_1}\,,\qquad A_2=R_2e^{iθ_2}\,,\qquad B=Se^{iθ}\,,
}
where $R_1$, $R_2$, $S$, $θ_1$, $θ_2$ and $θ$ are real. For the fixed points of the amplitude equations we set $\dot{A}_1=\dot{A}_2=\dot{B}=0$. For the existence of the fixed points, equations \eqref{eq22} and their complex conjugate pairs require that,
\equ{θ -θ_1 -θ_2=2nπ, \qquad n=0,±1,±2,⋯.
}
The equations for fixed points thus become,
\eqs{\big[ε -α\(R_1^2 +R_2^2\) -βS^2\big]R_1 +μR_2S &=0\,, \nn
\big[ε -α(R_1^2 +R_2^2) -βS^2\big]R_2 +μR_1S &=0\,, \nn
\big[ε -δ(R_1^2 + R_2^2) -γS^2]S +νR_1R_2 &=0\,,
}
which can be solved numerically to obtain the fixed points of the system. We perform the linear stability analysis on these equations to obtain the stabilities of the fixed points.
\begingroup
\allowdisplaybreaks
\eqs{s_1 &=\big[ε -α\(R_{10}^2 +R_{20}^2\) -βS_0^2\big] -\big[2α\(R_{10} +R_{20}\) +2βS_0\big]R_{10} +µ\(R_{20} +S_0\), \nn
s_2 &= \big[ε -α\(R_{10}^2 +R_{20}^2\) -βS_0^2\big] -\big[2α\(R_{10} +R_{20}\) +2βS_0\big]R_{20} +µ\(R_{10} +S_0\), \nn
s_3 &= \big[ε -δ\(R_{10}^2 +R_{20}^2\) -γS_0^2\big] -\big[2δ\(R_{10} +R_{20}\) +2γS_0\big]S_{0} +ν\(R_{10} +R_{20}\),
}
\endgroup
where $R_{10}$, $R_{20}$, and $S_0$ are the real parts of the complex amplitudes at fixed points.

Rhombic patterns that can be generated by the resonant triad \eqref{eq21} at a bicritical point are shown in Figure \ref{fig3}. For the $5:8$ forcing frequency in glycerol-water mixture the subharmonic tongue has a wavenumber $k_s=1.162$ and the harmonic tongue has a wavenumber $k_h=2.154$ at the bicritical point with an angle $22.02°$ between them. The triad amplitudes, that are used to generate the patterns, are the stable fixed points of the amplitude equations \eqref{eq22} calculated numerically for the coefficients $ε=0.301$, $α=0.5$, $β=0.1$, $γ=0.5$, $δ=0.1$, $μ=0.1$ and $ν=0.01$.
\fig{!h}{
\begin{subfigure}[t]{0.49\textwidth}
\caption{$φ=45°$}
\includegraphics[scale=0.55]{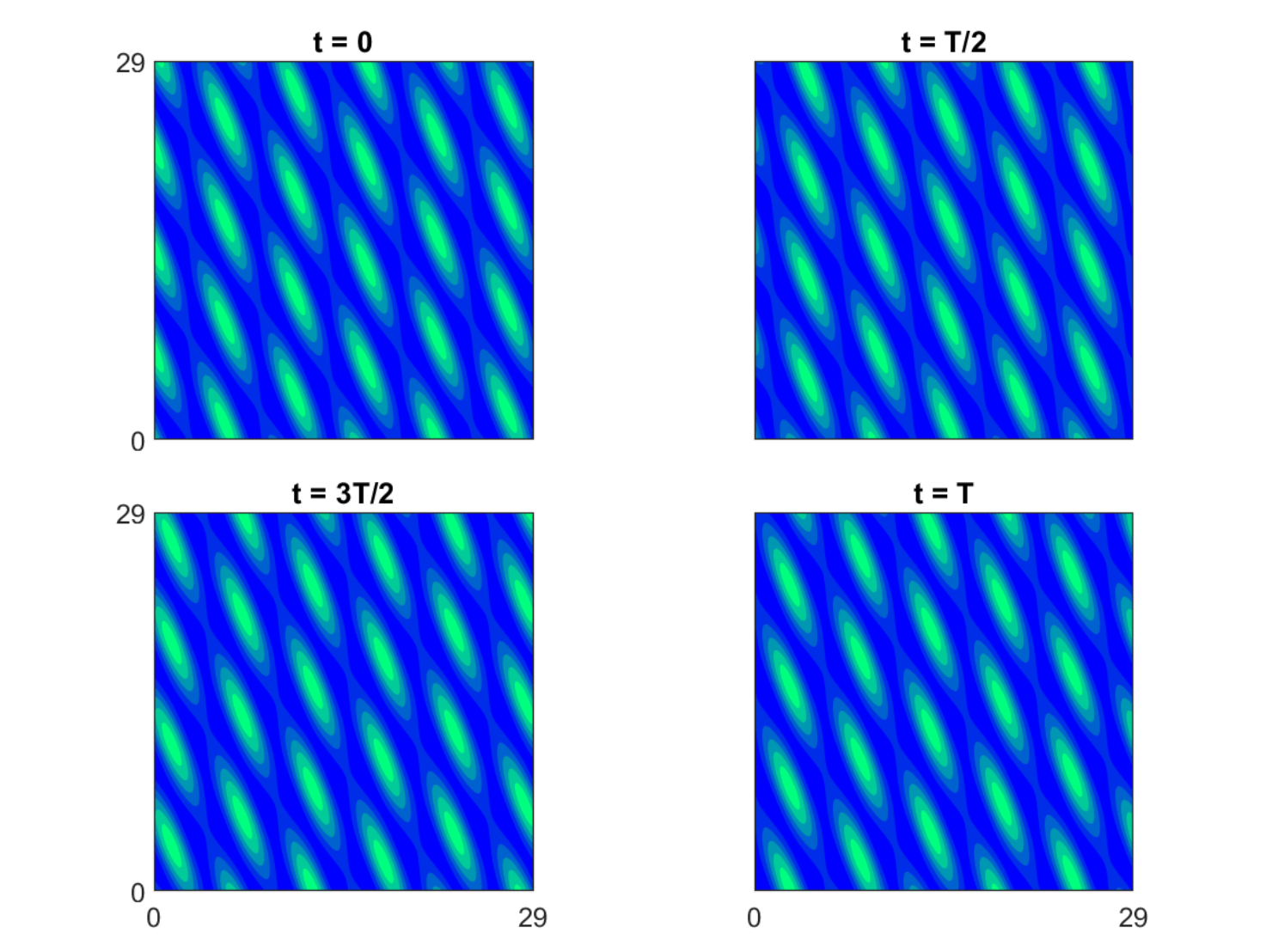}
\label{fig3a}
\end{subfigure}
\begin{subfigure}[t]{0.49\textwidth}
\caption{$φ=90°$}
\includegraphics[scale=0.55]{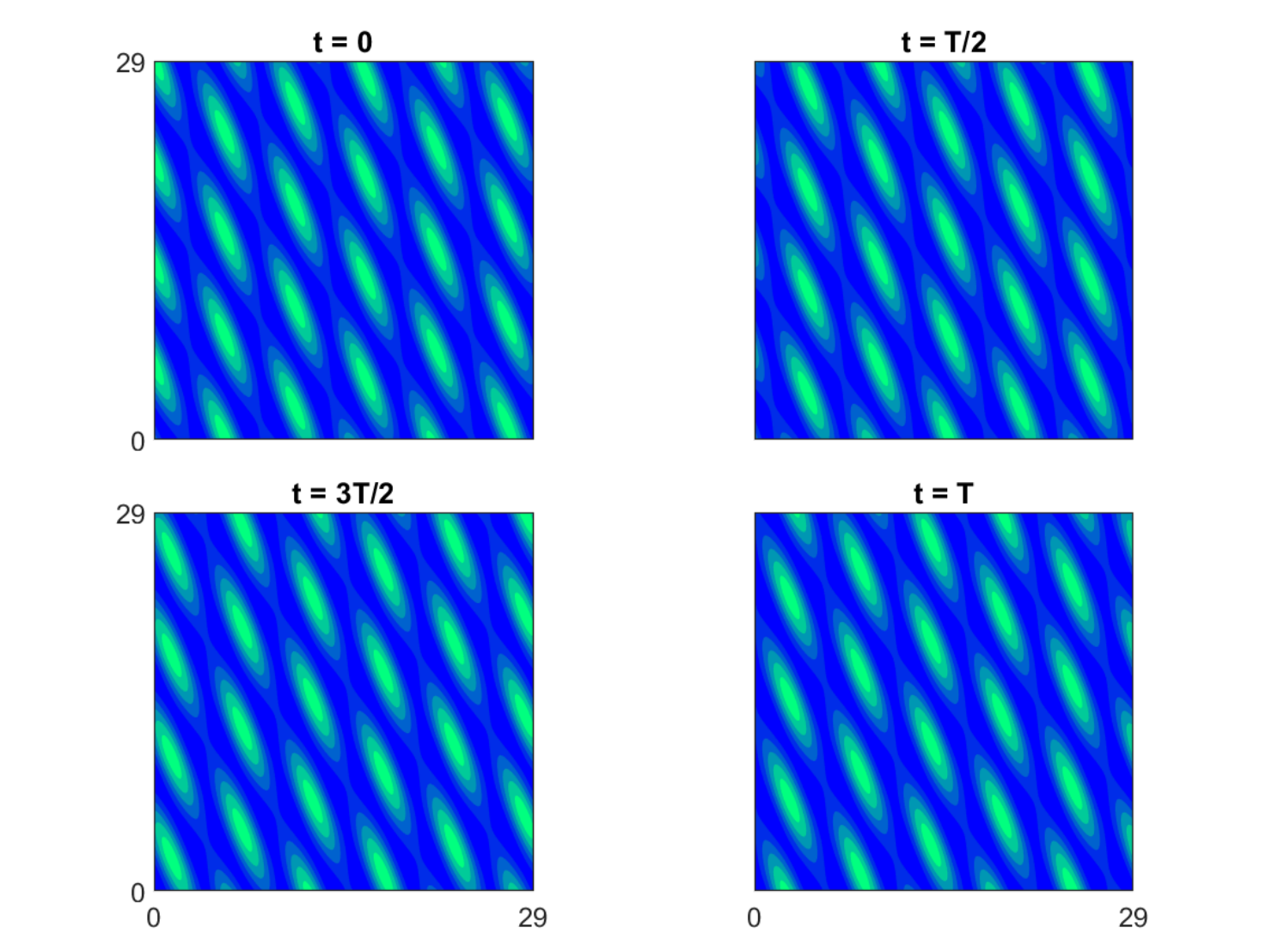}
\label{fig3b}
\end{subfigure}
\caption{Rhombic pattern with acute angle $44.04°$ at the primary instability as bicritical point for phase difference (a) $45°$ and (b) $90°$ with $A_1=A_2=0.5676$ and $B=0.6945$.}
\label{fig3}}

Another rhombic pattern is shown in Figure \ref{fig4} for the $21:34$ forcing in glycerol-water mixture. At the bicritical point the subharmonic wavenumber is $k_s=4.623$ and harmonic wavenumber is $k_h=8.281$; the two wavevectors are aligned with each other at an angle $26.48°$. The triad amplitudes are for $ε=0.120$, $α=0.5$, $β=0.1$, $γ=0.1$, $δ=0.1$, $μ=0.1$, $ν=0.1$.
\fig{!h}{
\begin{subfigure}[t]{0.49\textwidth}
\caption{$φ=45°$}
\includegraphics[scale=0.55]{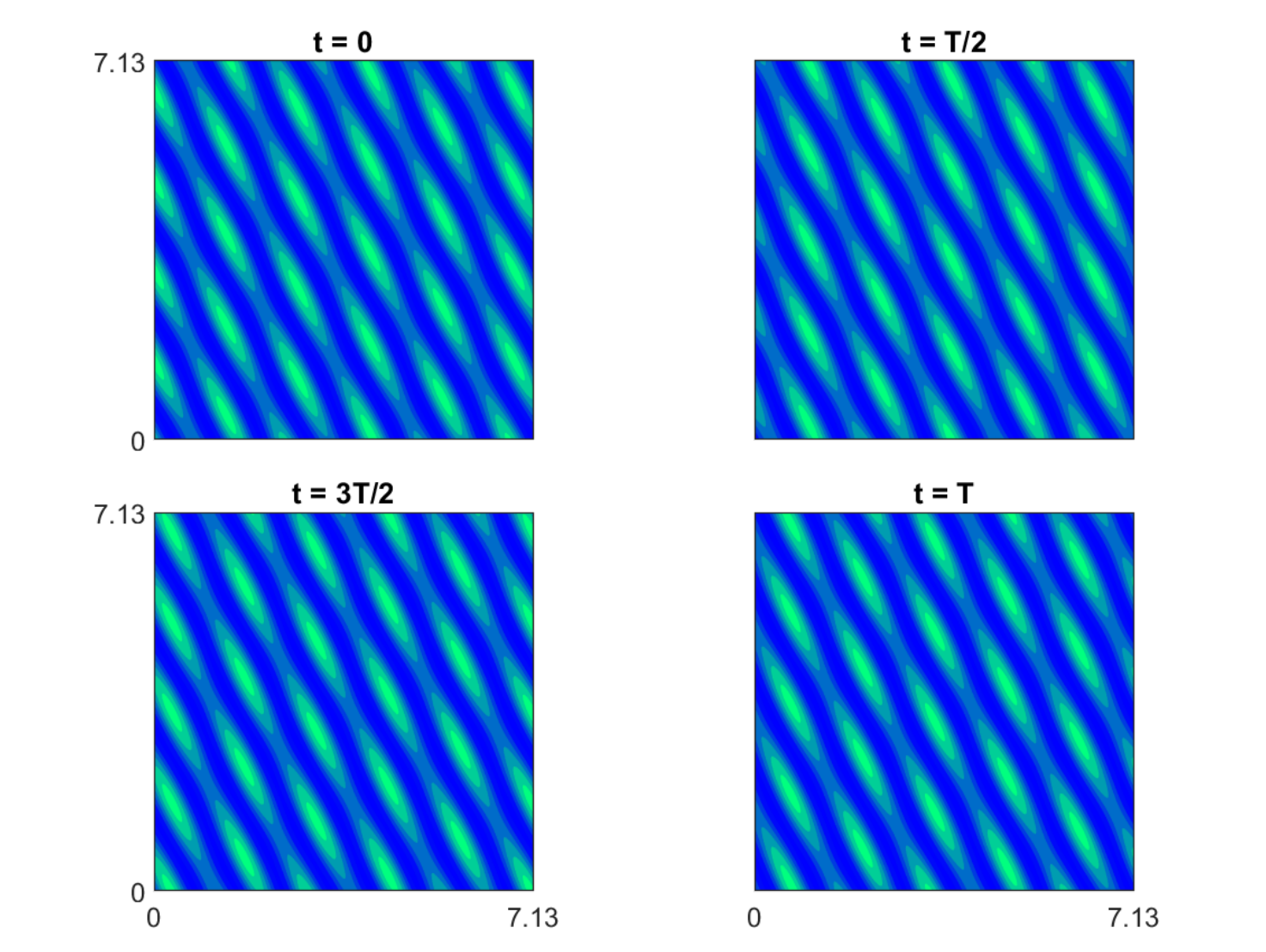}
\label{fig4a}
\end{subfigure}
\begin{subfigure}[t]{0.49\textwidth}
\caption{$φ=90°$}
\includegraphics[scale=0.55]{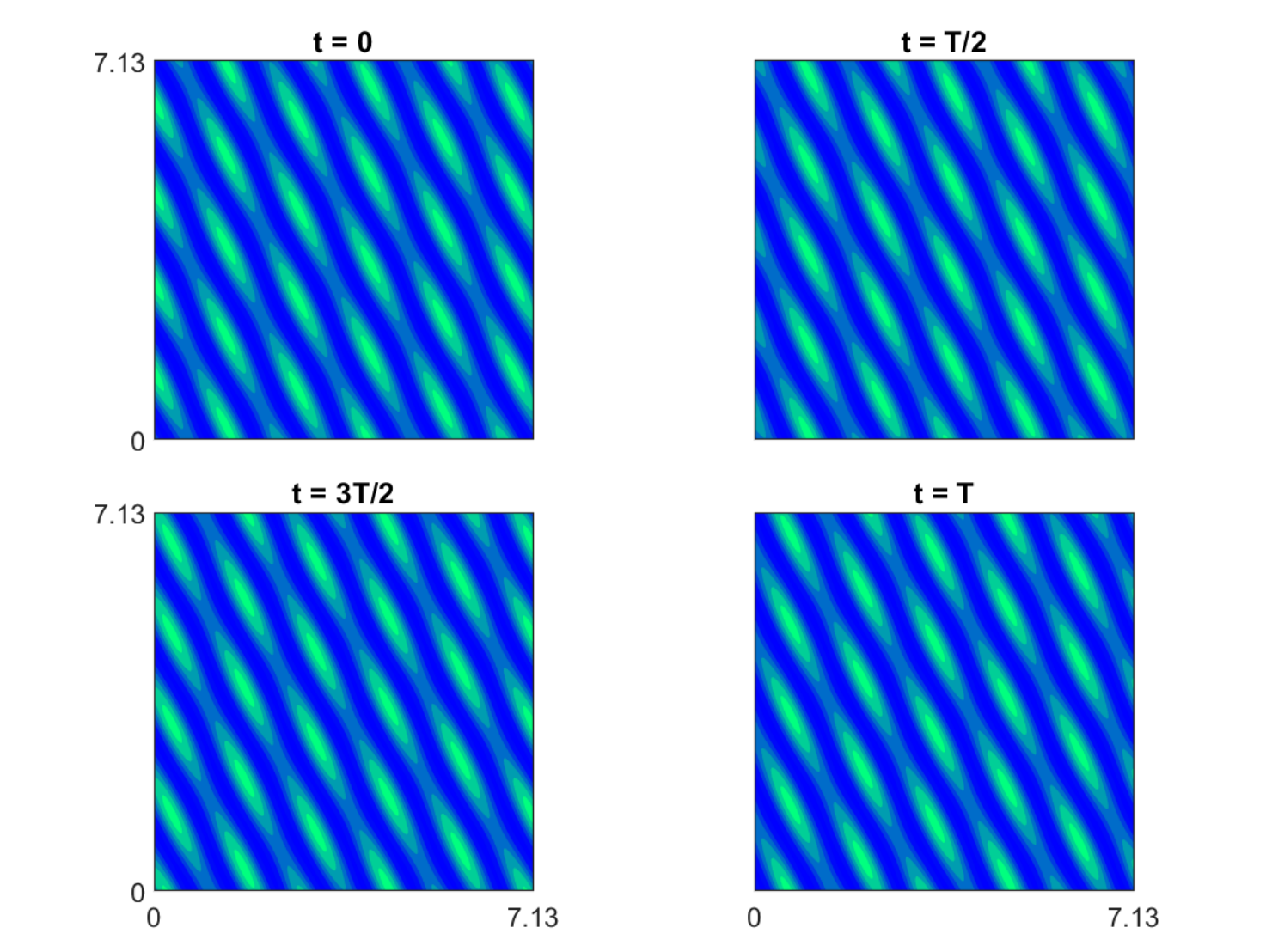}
\label{fig4b}
\end{subfigure}
\caption{Rhombic pattern with acute angle $52.96°$ at the primary instability as bicritical point for phase difference (a) $45°$ and (b) $90°$ with $A_1=A_2=0.3405$ and $B=1.0391$.}
\label{fig4}}

\section{Conclusion}
In this paper we have presented a systematic investigation of the effect of different frequency mixing in the parametric generation of surface waves by two frequency forcing. The frequency ratios which are used have nearly the same value but the response seems to depend on the actual frequencies used. For odd $l$ and $m$, we have a bicritical point involving $i^{th}$ and $j^{th}$ subharmonic tongues with $i=(l+1)/2$ and $j=(m+1)/2$. If the integer $l$ is even and $m$ odd, the $i^{th}$ harmonic tongue with $i=l/2$ and $j^{th}$ subharmonic tongue with $j=(m+1)/2$ are excited. We can determine which two solutions of the system will form the bicritical point for a given set of frequencies. This rule works for all kinds of situations, except for very thin layers of viscous fluids at low forcing frequency. We have proposed way of selecting a particular bicritical point \emph{a priori} at primary instability. The individual frequencies $lω$ and $mω$ decides the threshold of the bicritical points as primary instability. Our analysis would help us in choosing a desired pair of two possible wave numbers with predetermined temporal response. This may be useful in either exciting any desired pattern (\eg, superlattices or other exotic patterns) or analyzing any complex pattern at the instability onset. We have also shown how certain surface patterns can be formed by the competing wavenumbers at a bicritical point. The wavenumbers and the angle between the wavevectors (\ie, angle between $k_s$ and $k_h$) decide the shape of the periodic pattern. The nonlinear interaction between different modes play a major role in the selection of interesting and exotic patterns. Rhombic patterns appear naturally at the onset of bicritical points. We believe they should be easily accessible in experiments. The study can be used in generating several interesting and exotic patterns at the onset of parametrically driven systems.

\references{fluidrefs}

\end{document}